# Single crystal growth of TIMETAL LCB titanium alloy by floating zone method

J. Šmilauerová[a] *, J. Pospíšil[b], P. Harcuba[a], V. Holý[b], M. Janeček[a]

[a] *Faculty of Mathematics and Physics, Department of Physics of Materials, Charles University, Ke Karlovu 5, 121 16 Prague 2, Czech Republic*
[b] *Faculty of Mathematics and Physics, Department of Condensed Matter Physics, Charles University, Ke Karlovu 5, 121 16 Prague 2, Czech Republic*

\* Corresponding author
  E-mail address: smilauerova@karlov.mff.cuni.cz
  Tel.: +420221911614

**Abstract**
The methodology of single crystal growth of metastable β-Ti alloy TIMETAL LCB in an optical floating zone furnace is presented in this paper. Chemical compositions of both precursor material and single crystals were checked. It was found that the concentration of base alloying elements did not change significantly during the growth process, while the concentrations of interstitial elements O and N increased. DSC measurement determined that this concentration shift has a slight impact on ongoing phase transformations, as in the single-crystalline material peak associated with α phase precipitation moves by a few degrees to a lower temperature and peak attributed to diffusion controlled growth of ω particles shifts to a higher temperature. X-ray reciprocal space maps were measured and their simulation showed that the single crystal has a mosaic structure with mean size of mosaic blocks of approximately 60 nm.

## 1. Introduction

At room temperature and standard pressure, pure titanium crystallizes in a hexagonal close-packed (hcp) structure, which is known as the α phase. This phase is stable up to 883°C. Above this temperature the structure transforms to a body-centered cubic (bcc) β phase. The stability ranges of α and β phases can be altered and two phase regions or even new phases can be introduced by adding alloying elements. Titanium alloys are attractive materials for aerospace, automotive and biomedical applications due to their outstanding mechanical properties such as high strength in combination with low density. Particularly metastable β-Ti alloys remain of significant interest because of their excellent corrosion resistance, toughness and good hardenability through ageing treatment [1]. Metastable β-Ti alloys have enough β stabilizer content to retain the high temperature β phase in a metastable state upon quenching. In other words, the martensitic decomposition to the low temperature hexagonal α phase is prevented [2]. Moreover, metastable ω phase with hexagonal structure is formed in metastable β-Ti alloys with certain content of alloying elements. The ω phase is observed as uniformly dispersed submicron particles which are coherent with the β-Ti matrix. The ω phase is formed during quenching by a diffusionless displacive transformation, which was first described in detail by de Fontaine et al. [3]. The particles of ω phase further evolve and grow during ageing. This process is accompanied by rejection of β stabilizing elements from the ω phase; thus this reaction is diffusion controlled. The ω phase particles have an important influence on mechanical properties of the alloys. Typically, they increase specific strength and hardness but they can also embrittle the material [4]. Furthermore, the ω particles play a significant role in the β -> α phase transformations during ageing at higher temperatures, acting as nucleation sites for the α phase. Consequently, the resulting distribution, size and volume fraction of the α phase are determined by the preceding ω phase morphology [5, 6].

Despite a few decades of intensive research of the characteristics of ω phase particles, there is still doubt about the actual causes of ω phase formation. Some authors report that the initial step in ω formation is spinodal decomposition of β phase [7,8]. Other works also report elastic and structural instabilities as causes for ω particles formation during quenching [9, 10].

Since the mechanical properties of metastable β-Ti alloys depend strongly on the type and morphology of the particles of secondary phases (i.e. α and ω), the understanding of ongoing phase transformations, ageing kinetics and their influence on the resulting microstructure is of primary importance. Some experimental methods which study these aspects require single-crystalline material with known orientation. In particular, the topotaxial relation of the ω and β phases (i.e., mutual orientation of their lattices) can be studied by x-ray reciprocal space-mapping only in single crystals.

Small-angle x-ray scattering measurements of polycrystalline samples indicated a possible self-ordering of the ω-particles in the β matrix, however for a detailed analysis of the self-ordering, single-crystalline samples are required.

The objective of this work is to verify the feasibility of the growth of TIMETAL LCB (Ti - 6.8 Mo - 4.5 Fe - 1.5 Al in wt. %) single crystals by floating zone method in an optical furnace and to characterize the structural properties of these single crystals. TIMETAL LCB ("low cost beta") is an alloy which was designed primarily for application in automotive industry, e.g. suspension springs, engine valve springs and torsion bars [11]. The aim of its development was to produce a low cost alloy by means of selecting less expensive raw materials than typical β titanium alloys (i.e. less expensive ferromolybdenum master alloy is used) [12].

As TIMETAL LCB is a complex alloy containing three alloying elements, the main concern in single crystal growth is the melting and solidification behavior of the alloy. Fe belongs to a group of β eutectoid forming elements which typically cause solidification over a wider temperature range. This may lead to solute segregation during solidification of the material. On the other hand, Mo classifies as a β isomorphous alloying element. Addition of these elements does not result in wider solidification range and they are much less prone to segregation than β eutectoid elements. Al is an α stabilizing element and is known to form Al-rich regions in some cases [2]. Nevertheless, the segregation can be eliminated by homogenization (i.e. solution treatment) of the ingot. To the authors' knowledge, no incongruent melting was reported in metastable β titanium alloys.

The chemical compositions of precursor and grown ingot were checked and compared. Furthermore, the effect of compositional change during crystal growth on phase transformations was studied by differential scanning calorimetry (DSC).

The grown ingots were used for x-ray diffraction studies which investigated the influence of ageing on the evolution of ω particles and examined their lattice misfit. The results of these experiments are also discussed in this paper.

According to literature, only a few successful attempts to grow metastable β-Ti alloys using floating zone method were made and published. Lee et al. [13] studied single-crystals of Ti-15Mo-5Zr-3Al alloy and their plastic deformation behavior. Takesue et al. [14] investigated the growth and elastic properties of Ti–Nb–Ta–Zr–O alloy (Gum Metal). Hermann et al. [15] reported the growth and characterization of Ti-45Nb single-crystals which were grown using two-phase radio-frequency heating.

## 2. Experimental details

In this research, we have attempted to grow a single crystal of a metastable β-Ti alloy, TIMETAL LCB. Polycrystalline rods (diameter of 8 mm and length typically 100 mm) were used as precursors for growth of the crystals. The rods were prepared by arc melting by TIMET Corporation. A series of single crystals of the TIMETAL LCB alloy has been prepared by floating zone method in a commercial four-mirror optical furnace with halogen lamps 4×1000 W (model FZ-T-4000-VPM-PC, Crystal Systems Corp., Japan). After the growth process, each crystal was solution treated at 860°C for 4 h in an evacuated quartz tube and subsequently water quenched. This treatment ensured homogenized structure and retention of metastable β phase. The quality of all grown crystals was tested by Laue method using a dedicated Laue camera equipped with a low-power microfocus source (W white x-ray radiation) and a large two-dimensional solid-state detector. We used the OrientExpress software [16] to determine the orientation of the single crystals.

In order to estimate the compositional changes in the material during crystal growth process, chemical compositions of both precursor and single crystals were measured. Two complementary techniques were employed. The content of the main alloying elements (Ti, Mo, Fe and Al) was determined by energy dispersive x-ray spectroscopy (EDS) on scanning electron microscope FEI Quanta 200FEG using software Genesis by EDAX. As titanium and its alloys are prone to interstitial contamination at elevated temperatures, especially by oxygen and nitrogen, the content of these two elements was analyzed using an automatic analyzer LECO TC 500C.

The EBSD analysis was performed on scanning electron microscope FEI Quanta 200FEG using DigiView EBSD detector to confirm macroscopic crystallinity of the grown ingots. The analysis of the data was carried out using OIM software.

Differential scanning calorimetry (DSC) runs were performed on Netzsch calorimeter DSC 404 Pegasus. During the measurement, the sample was held in an inert Ar atmosphere with flow rate of 30 ml/min.

In order to obtain samples of appropriate size and shapes for the above mentioned experimental methods, the single crystals were cut using an automatic precision saw Struers Accutom-50 with Struers wafering blade. The samples were then ground and polished utilizing 500, 800, 1200, 2400 and 4000 grit SiC papers. Final polishing was done using a vibratory polisher and employing 0.3 μm and 0.05 μm aqueous alumina ($Al_2O_3$) suspensions and 0.05 μm colloidal silica.

## 3. Details of the single crystal growth process

The quartz chamber of the optical furnace was evacuated by a turbomolecular pump up to $10^{-6}$ mbar before each crystal growth process to avoid oxidation of the titanium alloy. In order to desorb possible surface oxygen contamination on the precursor, the furnace power was increased gradually up to 30% of maximum power (far from the melting point of the alloy) and the rod precursor was passed through the hot zone while continuously evacuating. Significant deterioration of the vacuum down to $10^{-3}$ mbar was observed when the precursor warmed up, which indicated desorption of oxygen and other gases from the surface of the precursor. After proper evacuation the quartz chamber was quickly filled with high purity Ar protective atmosphere (6N). The whole growth process was performed with Ar flow of 0.25 l/min and pressure of 2.5 bar. At the beginning of each crystal growth, a neck was created in order to isolate one grain. The pulling rate was 10 mm/h and the rotation velocity of both the upper and the lower shaft was 5 rpm. Throughout the whole crystal growth process, the molten zone was perfectly clear with no signs of oxides or foreign phases on the surface, manifesting very high quality of the precursor.

However, some unusual effects have been observed during the growth of the single crystals. A part of the ingot (up to 10 mm below the hot zone) seemed to be still coated with a thin layer of liquid which solidified unevenly on the surface of the ingot. This resulted in irregularities in cylindrical shape of the crystal ingot (see Fig. 1). Initially we suspected that the hot zone was overheated. Nevertheless, when the power of the furnace was reduced, the effect remained almost unchanged while the upper and the lower rod came into contact, as evidenced by off-centered rotation of the rods and vibrations of the melt. The effect of overheating may be therefore excluded. Another possibility which can be considered is very low thermal conductivity of titanium in comparison to other metals. The poor heat transfer through the ingot could be responsible for surface melting due to the absorption of the scattered light. Another hypothesis which could explain the occurrence of the liquid layer on the surface of the alloy is incongruent melting. However, we did not find any spurious phases inside the crystal which often point to incongruently melting materials. A possible explanation may be that the liquid layer results from the combination of slight compositional changes on the surface of the melt and flow patterns in the liquid zone. Currently, we do not have any unambiguous experimental evidence to decide what causes the observed effect.

The total length of grown crystals was up to 9 cm, one of the grown ingots is shown in Fig. 1. The beginning of the single crystal near the neck exhibits coloring which indicates that the surface layer is contaminated by interstitial elements. With increasing distance from the neck, the coloring gradually disappears and the ingot exhibits clean metallic surface. This corresponds to the fact that the furnace and the precursor itself are releasing small amounts of gases at the beginning of single crystal growth even after proper degassing procedure described above.

## 4. Results

The quality and the orientation of all grown crystals were determined by Laue method. Two samples from each crystal - one from the beginning and the other from the end of the ingot - were cut and polished using methods described in Section 2. The spot size of the beam (0.5 mm) was much lower than the diameter of the crystal, which allowed us to record a series of Laue patterns at different points on the surface of each sample. Laue patterns from the two samples cut from the same crystal ingot exhibited the same symmetry of the diffraction spots, which indicates single grain within the whole length. All observed diffraction patterns exhibited sharp spots which refer to high quality of the single crystal. One representative Laue pattern which was recorded in central part of one of the slices is displayed in Fig. 2. Laue pattern analysis confirmed that the structure of our crystal corresponds to β phase. The growth direction of the single crystals was approximately <521>.

Selected samples from different sections of each of the grown crystals were also checked using EBSD analysis. It was found that the grown crystals consisted of one central single-crystalline grain which extended over the whole length of the ingot. This central grain was sometimes covered by a thin layer of smaller grains with maximum thickness of 1-2 mm (see Fig. 3). The layer usually consisted of only few grains and the thickness of the layer varied over the length of the crystal ingot. Nevertheless, the central grain usually filled up most of the grown ingot cross-section. We suppose that the grains on the circumference of the ingot arise from the liquid layer observed in the vicinity of the hot zone during the growth process. The presence of the liquid layer was discussed in Section 3.

From the results of Laue diffraction and EBSD analysis it was concluded that the crystallinity of all grown crystals had essentially the same character. One dominant central single-crystalline grain was present in all ingots. The analysis of Laue diffraction patterns proved good quality of all grown crystals. Therefore, it can be stated that all the resulting ingots contain sufficient volume of single crystal grain and the process of single-crystal growth is well reproducible.

The structure quality of the single crystals was studied in detail by x-ray diffraction reciprocal-space mapping technique [17]. We used a laboratory x-ray diffractometer (CuK$\alpha$ radiation, 40 kV/35 mA) equipped with a parabolic multilayer optics, a parallel-plate collimator and a secondary graphite monochromator to suppress the fluorescence radiation. Figure 4 (a) presents the reciprocal-space distribution of the diffracted intensity (reciprocal-space map) measured around the $[112]_\beta$ diffraction in an asymmetric coplanar scattering geometry. We compared the shape of the diffraction maximum with the simulations based on the mosaic-block model [17], in which the crystal structure is modelled by randomly placed and randomly rotated mosaic blocks, the x-ray waves diffracted by different blocks are assumed incoherent. In the simulations we took into account the actual resolution function of the diffractometer. The best match of the measured and simulated data was obtained for spherical blocks with the diameter of (60 ± 5) nm and root-mean square misorientation smaller than 0.01 deg.

Important concern regarding single crystal growth of titanium alloys is the contamination of the material by interstitial elements and other compositional changes. Interstitial elements, most importantly O and N, are responsible for changes in mechanical properties [2, 18]. Moreover, phase transformations and their onsets can be altered with compositional changes in the alloy.

The chemical compositions of both precursor material and single crystal were determined by the two methods described in Section 2. The results are summarized in Table 1 for precursor material as well as for single crystal. EDX measurements were performed on three samples cut from different parts of the single crystal and the obtained values were averaged, as they did not shift significantly across the length of the crystal. However, the content of interstitial elements in the single crystal should be regarded as approximate, as it was determined from a small volume in the central part of the ingot. Nevertheless, slight shifts in concentration of N and O across the length of the single crystals are possible. The chemical analysis indicates that the concentration of base alloying elements (Ti, Mo, Fe, Al) did not change significantly during the growth process. The values for these elements given in Table 1 coincide within the error limits. On the other hand, the concentrations of both N and O increased during single crystal growth.

The shift in the amount of interstitial elements may influence the phase transformations occurring in the material. Therefore, DSC measurements were done on the precursor material and the single crystal in order to estimate these changes in phase transformations. Two samples from single-crystalline material were prepared in order to assess the change in phase transformations along the length of the ingot. One single-crystalline sample was taken from the part near the neck and another sample was cut near the end of the single crystal. The comparison of the DSC runs is given in Fig. 5. The heating rate was 5 °C/min and the maximum temperature was 650 °C. Certain differences in DSC curves can be observed. For example, the sharp peak near 500 °C which is attributed to α phase precipitation (see e.g. [19, 20]) is shifted to a lower temperature for single crystal measurement. In the precursor, the maximum of this peak is found at 509 °C. Single crystal DSC curve from the sample near the neck exhibits the maximum of this peak at 501 °C. The peak for the sample from the end of the single crystal is shifted to approximately 502 °C. The smaller peak between 300 °C and 400 °C corresponding to diffusion controlled growth of ω particles [19, 20] shows a tendency of shifting to a higher temperature after the growth process. The position of the maximum is 345 °C for the precursor. In the single crystal DSC curve, this peak is shifted to 379 °C for the sample near the neck of the single crystal and to 375 °C for the sample taken from the end of the single crystal. These changes in phase transformations can be explained by an increased content of interstitial elements in the single-crystalline material. It is well-known that oxygen is an α stabilizing element [2]. Therefore, higher concentration of oxygen in single crystal shifts the sharp α peak to lower temperatures. Moreover, increased O content probably hinders the growth of ω particles which results in the ω peak being found at a higher temperature for the single crystal when compared to the precursor material. From the comparison of peak positions on both single crystal DSC measurements (sample near the neck and sample near the end of the single crystal) it can be inferred that the part near the beginning of the single crystal contains more O and N impurities. This is consistent with the fact that at the beginning of single crystal growth, some amount of gases is released by the furnace and the precursor even after careful degassing procedure.

## 5. Summary

The growth of single crystals of one of the metastable β-Ti alloys, TIMETAL LCB, was successfully accomplished. The floating zone method proved to be a suitable method for single crystal growth of this type of material. We have determined the crystal structure and quality of the grown crystals by Laue diffraction and EBSD analysis. These experimental techniques confirmed high quality of the obtained single crystals. The chemical composition change during single crystal growth proved that the concentrations of interstitial N and O increased. On the other hand, the concentrations of the main alloying elements (Ti, Mo, Fe and Al) did not change significantly during the growth process. DSC measurements indicated that the phase transformations in the material are slightly altered due to increased concentration of interstitial elements. Peak in DSC curve attributed to the growth of ω particles shifted to a higher temperature. The second peak corresponding to α phase precipitation moved to a lower temperature. Using reciprocal space mapping method we studied the mosaic structure of the single crystals and we found that the mean size of the mosaic blocks is about 60 nm.

**Acknowledgements**

This work was financially supported by the Czech Science Foundation under the project P107-12-1025. Partial financial support was provided by Grant Agency of Charles University (GAUK) - 106-10/251403. Single crystal growth experiments were performed in MLTL (http://mltl.eu/), which is supported within the program of Czech Research Infrastructures (project no. LM2011025). The authors would like to acknowledge the Clemson University, SC, USA, namely prof. H. J. Rack, for providing the precursor material.

**References**
[1] P. J. Bania, Journal of the Minerals, Metals and Materials Society, 46(7): 16-19, 1994.
[2] G. Lütjering and J.C. Williams, *Titanium*, Springer-Verlag, 2007.
[3] D. De Fontaine, N. E. Paton, and J. C. Williams, Acta Metallurgica, 19(11): 1153-1162, 1971.
[4] T. Sakamoto, K. Nakai, M. Maeda and S. Kobayashi, Materials Science Forum, 561: 2067-2070, 2007.
[5] J. C. Williams, Titanium Science and Technology, 1433-1494, 1973
[6] F. Prima, P. Vermaut, G. Texier, D. Ansel and T. Gloriant, Scripta Materialia, 54(4): 645-648, 2006.
[7] M. K. Koul and J. F. Breedis, Acta Metallurgica, 18(6): 579 - 588, 1970.
[8] H. P. Ng, A. Devaraj, S. Nag, C. J. Bettles, M. Gibson, H. L. Fraser, B. C. Muddle, R. Banerjee, Acta Materialia, 59(8): 2981 - 2991, 2011.
[9] D. de Fontaine, Acta Metallurgica, 18(2): 275 - 279, 1970.
[10] A. Devaraj, S. Nag, R. Srinivasan, R. E. A. Williams, S. Banerjee, R. Banerjee and H. L. Fraser, Acta Materialia, 60(2): 596 - 609, 2012.
[11] S. L. Nyakana, J. C. Fanning, and R. R. Boyer, Journal of Materials Engineering and Performance 14(6): 799 - 811, 2005.
[12] P. G. Allen, P. J. Bania, A. J. Hutt, and Y. Combres, Titanium '95 Science and Technology, 1680-1687, 1995.
[13] S. H. Lee, K. Hagihara, M. H. Oh, and T. Nakano, Journal of Physics: Conference Series, Vol. 165, No. 1, p. 012086, IOP Publishing, 2009.
[14] N. Takesue, Y. Shimizu, T. Yano, M. Hara, and S. Kuramoto, Journal of Crystal Growth 311(12): 3319-3324, 2009.
[15] R. Hermann, M. Uhlemann, H. Wendrock, G. Gerbeth, and B. Büchner, Journal of Crystal Growth 318(1): 1048-1052, 2011.
[16] J. Laugier and B. Bochu, Orient Express Version 3.4, Laboratoire des Materiaux et du Génie Physique de l'Ecole Supérieure de Physique de Grenoble, 2005.
[17] U. Pietsch, V. Holý, and T. Baumbach, *High-Resolution x-ray Scattering From Thin Films to Lateral Nanostructures*, Springer-Verlag Berlin, Heidelberg, New York 2004.
[18] Ye. V. Petunina, Metal Science and Heat Treatment of Metals, 3(5-6): 276 - 279, 1961.
[19] F. Prima, P. Vermaut, D. Ansel, J. Debuigne, Materials Transactions - JIM, 41(8): 1092 - 1097, 2000.
[20] Q. Contrepois, M. Carton, J. Lecomte-Beckers, Open Journal of Metal, 1, 1 - 11, 2011.

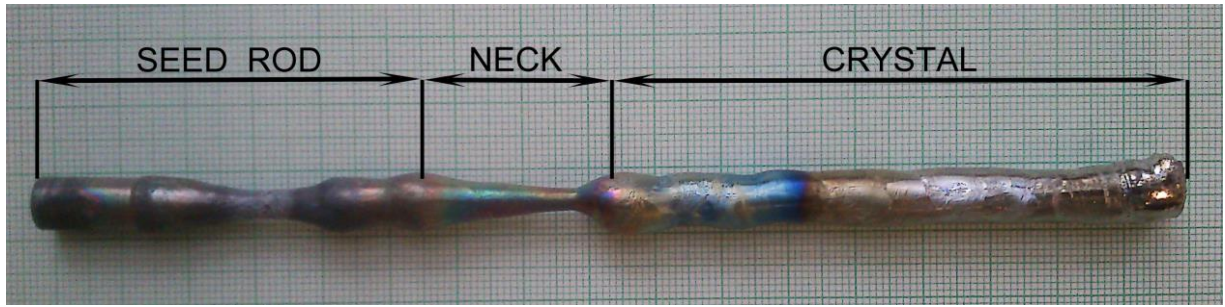

**Fig. 1.** (Color online) Example of a single crystal grown in the optical floating zone furnace. Seed rod, neck and single crystal are indicated.

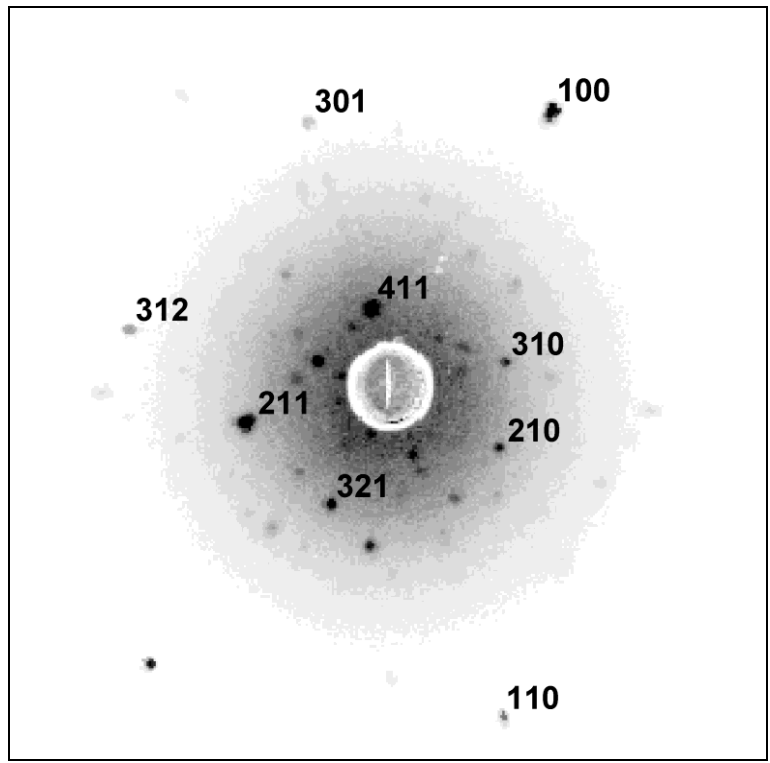

**Fig. 2.** A representative Laue pattern of TIMETAL LCB single crystal. Sharp and symmetrical spots confirm high quality single crystal.

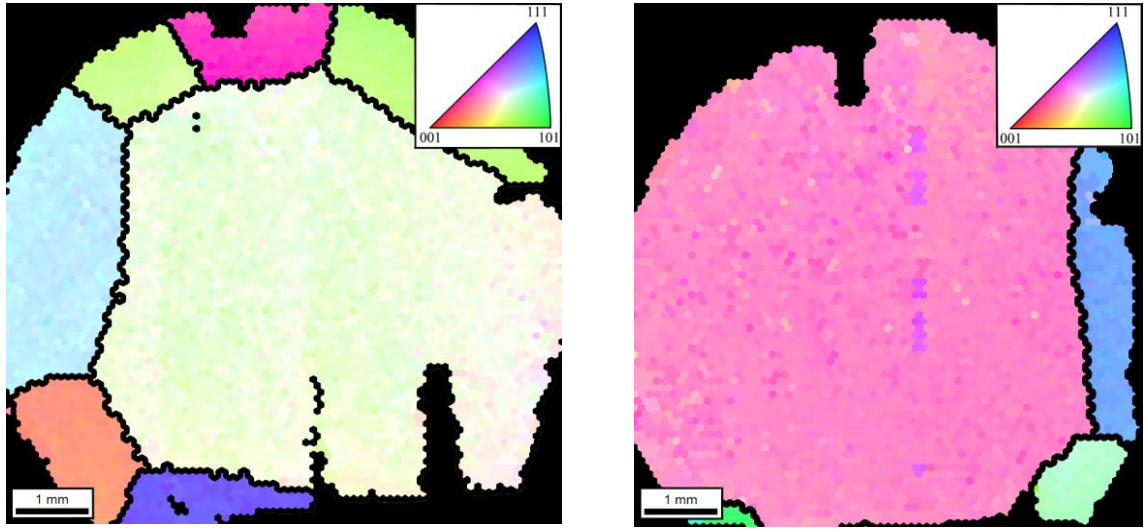

**Fig. 3.** (Color online) EBSD map measured on the cross-sections of two different crystals grown by the method described in this paper. Both patterns show a large central single crystal grain which extends over the whole length of each of the crystal ingots. In the left panel, a polycrystalline layer composed of a few large grains is evident around the central single crystal grain. In the right panel, almost single-crystalline cross-section of one of the ingots is displayed.

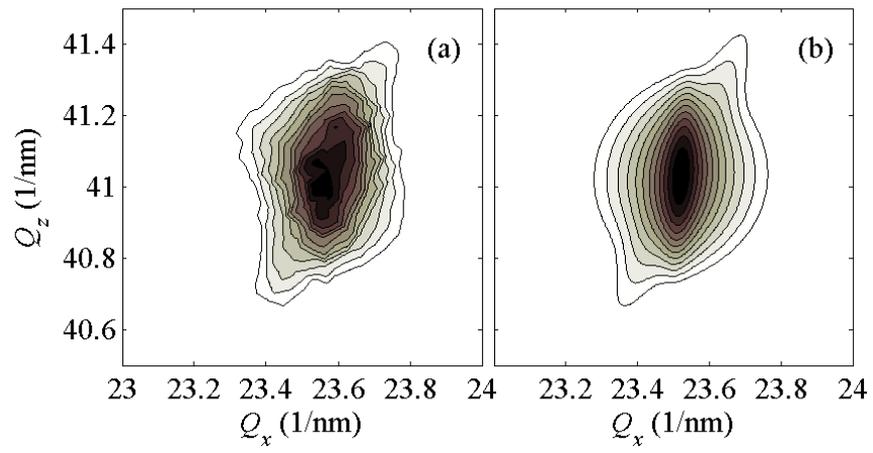

**Fig. 4.** Measured (a) and simulated (b) diffraction reciprocal space maps in diffraction $[112]_\beta$. The step of the iso-intensity contours is $10^{0.2}$.

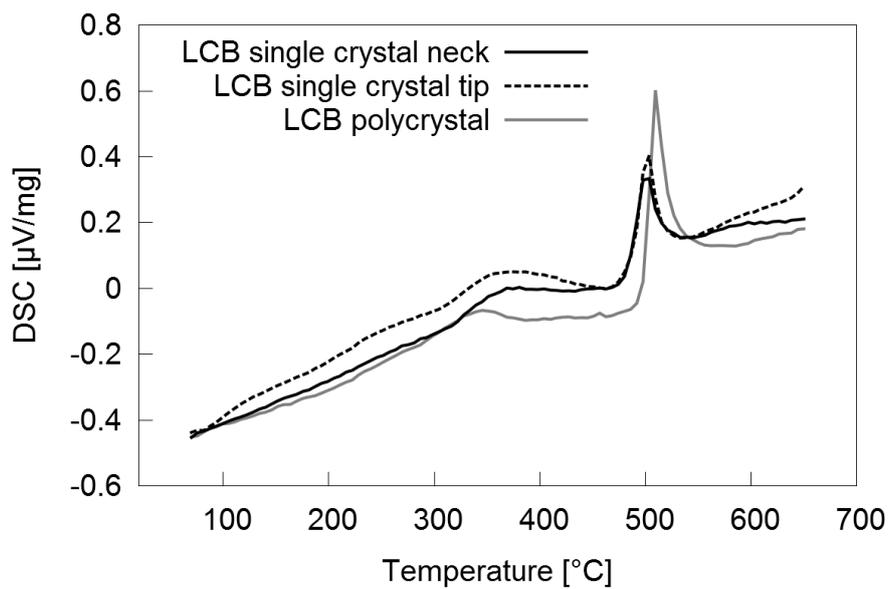

**Fig. 5.** DSC curves for precursor material (grey) and single crystal (black). DSC measured on a sample from the part of the single crystal near the neck is denoted by a solid line; DSC on a sample near the end of the single crystal is denoted by a dashed line. The heating rate was 5 °C/min for all runs. Exothermic peaks point up.

**Tab. 1.** Chemical composition of initial material (TIMETAL LCB rods) and grown single crystal in at.%. Given values are averages and their standard deviations calculated from three measurements on polycrystalline precursors. Ti, Mo, Fe and Al compositions were determined by EDX; compositions of N and O were determined by analyzer LECO TC 500C. As the chemical composition was determined by two independent methods, the sum of at.% is not 100%.

| Element | Ti | Mo | Fe | Al | N | O |
|---|---|---|---|---|---|---|
| Precursor | 88.7 ± 0.7 | 4.1 ± 0.4 | 3.7 ± 0.5 | 3.1 ± 0.2 | 0.011±0.004 | 0.40±0.06 |
| Single crystal | 88.4 ± 0.4 | 4.3 ± 0.3 | 3.4 ± 0.4 | 3.1 ± 0.2 | 0.29±0.03 | 0.55±0.06 |